\documentclass[acmtog]{acmart}
\usepackage{array}
\usepackage{makecell}
\usepackage{wrapfig}
\usepackage{tcolorbox}
\usepackage{booktabs}   
\usepackage{multirow}   

\usepackage{duckuments}
\usepackage{tikzducks}
\usepackage[capitalise]{cleveref}
\usepackage{svg}
\usepackage{graphicx}
\usepackage{tikz}
\usepackage{float} 
\usepackage{animate}
\usepackage{bm}
\usepackage[linesnumbered,ruled,vlined]{algorithm2e}
\usepackage{titlesec}
\usepackage{subcaption}

\titleformat{\paragraph}
{\normalfont\normalsize\bfseries}{\theparagraph}{1em}{}
\titlespacing*{\paragraph}
{0pt}{3.25ex plus 1ex minus .2ex}{1.5ex plus .2ex}

\usepackage{xcolor}

\SetCommentSty{mycommfont}

\SetKwRepeat{Do}{do}{while}

\definecolor{SithColor}{rgb}{0.7,0,0} 
\definecolor{ConsularColor}{rgb}{0,0.4,0} 
\definecolor{GuardianColor}{rgb}{0,0,0.8} 
\definecolor{WinduColor}{rgb}{0.56,0.34,0.62} 

\setcopyright{none}
\renewcommand\footnotetextcopyrightpermission[1]{}
\begin{document}

\title{HairFormer: Transformer-Based Dynamic Neural Hair Simulation}

\author{Joy Xiaoji Zhang}
\email{joyxiaojizhang@cs.cornell.edu}
\orcid{1234-5678-9012} 
\affiliation{%
  \institution{Cornell University}
  \city{Ithaca}
  \state{New York}
  \country{USA}
  \postcode{14850}
}

\author{Jingsen Zhu}
\email{jz2358@cornell.edu}
\affiliation{%
  \institution{Cornell University}
  \city{Ithaca}
  \state{New York}
  \country{USA}
  \postcode{14850}
}

\author{Hanyu Chen}
\email{hc2269@cornell.edu}
\affiliation{%
  \institution{Cornell University}
  \city{Ithaca}
  \state{New York}
  \country{USA}
  \postcode{14850}
}

\author{Steve Marschner}
\affiliation{%
  \institution{Cornell University}
  \city{Ithaca}
  \state{New York}
  \country{USA}
  \postcode{14850}
}
\email{srm2@cs.cornell.edu}
\begin{teaserfigure}
  \centering
  \includegraphics[width=\textwidth]{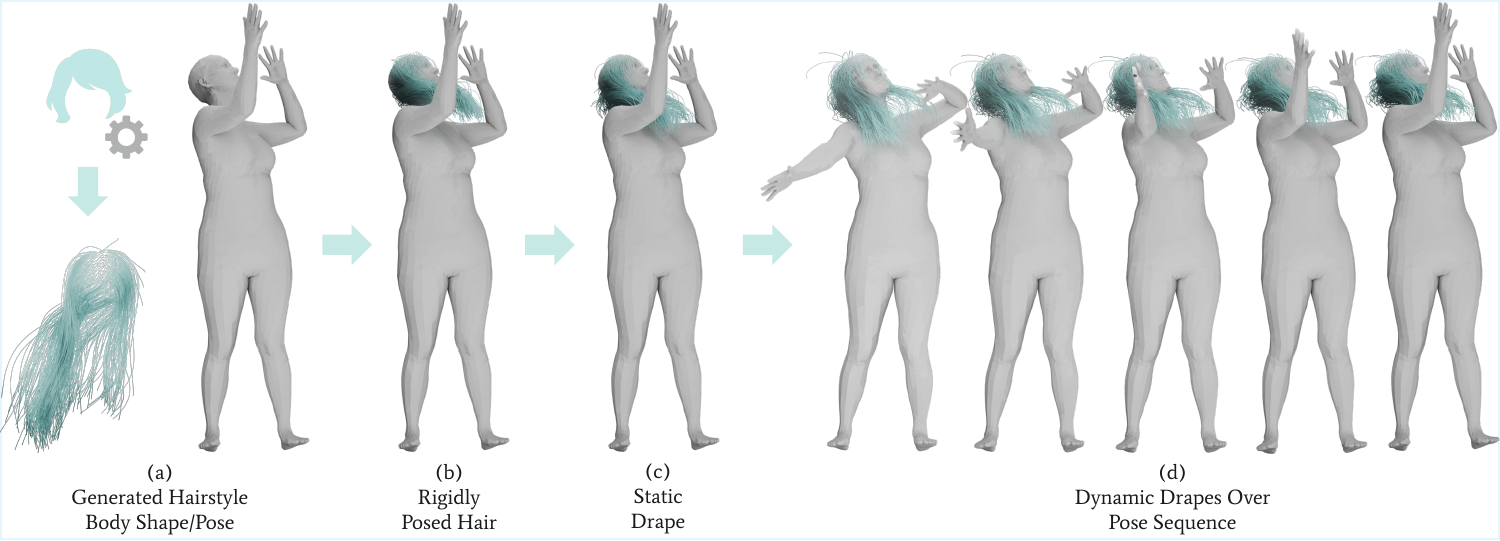}
  \caption{\textbf{The HairFormer pipeline for dynamic neural hair simulation:} (a) Given an arbitrarily generated hairstyle, body shape, and a body pose (or pose sequence), (b) the hair is first rigidly positioned on the body. (c) A Transformer-powered static network then predicts the static draped shape for single frames, effectively resolving hair-body penetrations.  (d) Finally, a dynamic network leverages the static predictions to generate expressive dynamic motions, introducing effects such as flying "antenna hairs" and enhanced overall hair movement, along with complex secondary effects over the full pose sequence. }
  \label{fig:teaser}
\end{teaserfigure}


\begin{abstract}
Simulating hair dynamics that generalize across arbitrary hairstyles, body shapes, and motions is a critical challenge. Our novel two-stage neural solution is the first to leverage Transformer-based architectures for such a broad generalization. We propose a Transformer-powered static network that predicts static draped shapes for any hairstyle, effectively resolving hair-body penetrations and preserving hair fidelity. Subsequently, a dynamic network with a novel cross-attention mechanism fuses static hair features with kinematic input to generate expressive dynamics and complex secondary motions. This dynamic network also allows for efficient fine-tuning of challenging motion sequences, such as abrupt head movements. Our method offers real-time inference for both static single-frame drapes and dynamic drapes over pose sequences. Our method demonstrates high-fidelity and generalizable dynamic hair across various styles, guided by physics-informed losses, and can resolve penetrations even for complex, unseen long hairstyles, highlighting its broad generalization. 
\end{abstract}




\maketitle
\section{Introduction}

Realistic 3D hair modeling and simulation is crucial for creating high-fidelity digital humans in applications like visual effects and virtual reality. However, accurately simulating hair dynamics in real-time remains a significant challenge, especially for unseen assets with diverse hairstyles, body poses, and shapes. Accurately capturing the complex geometry and dynamic behavior of individual strands in response to varied body movements is an active research area. While substantial progress has been made in modeling and generating diverse hairstyles, often using generative models like StyleGAN \cite{karras2019style} to produce a wide array of appearances \cite{he2024perm}, and in simulating static draped states \cite{stuyck2024quaffure}, it remains difficult to simulate hair dynamics that include complex secondary motions for arbitrary hairstyles on an animated body. Existing static or quasi-static approaches typically model hair as single-frame drapes and are limited by their lack of temporal context. This means they struggle to predict how individual hair strands react to dynamic forces, resulting in a "frozen" or "glued" appearance during motion, significantly hindering the realism and expressiveness of digital humans.

To address these limitations, this paper introduces a novel two-stage neural simulation technique that works for arbitrary hairstyles, body poses, and body shapes and produces a wide range of dynamic effects, including detailed secondary motions, for realistic hair animation that reflects character movement, all while enabling real-time inference for both static and dynamic stages.

Our method first employs a Transformer-based~\cite{vaswani2017attention} static network to establish a physically plausible static draped shape for any given hairstyle, even those unseen during training. This network takes as input a hair latent code (used by StyleGAN to generate initial hair strands), a single body pose, and body shape parameters. It encodes the hair and body information separately, then uses a cross-attention mechanism to capture the spatial relationships between the hair strands and the body. The network is trained in a self-supervised manner and explicitly resolves penetrations of hair strands into the body mesh. This is achieved using an extended barrier function, inspired by Incremental Potential Contact (IPC) \cite{li2020codimensional}, which is crucial to maintaining hair fidelity by preventing unnatural hair stretching and preserving the inherent characteristics of hairstyles, especially for unseen configurations.

Building upon this static prediction, our dynamic network integrates temporal information from a sequence of body poses. Its core architecture features a cross-attention module that fuses geometric details from the static draped hair with kinematic features (velocity and acceleration) from the motion sequence. This fusion mechanism is crucial as it allows the network to preserve the intricate details and characteristics of a static hairstyle while accurately predicting detailed dynamic responses, including complex secondary motions such as ``flying'' or ``antenna'' hair induced by drastic body movements. This robust architecture promotes coherence between static and dynamic states, and enables rapid fine-tuning (in only a few hundred iterations) to excel on challenging pose sequences, such as fast head rotations and abrupt stops, which significantly enhances realism in such demanding scenarios. The dynamic network is further guided by an inertia loss term and a Gated Recurrent Unit (GRU) \cite{chung2014empirical}, both contributing to simulated hair motions that are physically plausible and temporally coherent.

Although inspired by previous work on hair generation \cite{karras2019style, he2024perm}, neural cloth simulation \cite{bertiche2020pbns,bertiche2022neural}, and quasi-static hair \cite{stuyck2024quaffure}, our approach introduces crucial advances. To our knowledge, this work is the first to simulate generalized dynamics that includes complex secondary motions for arbitrary and unseen hairstyles that interact with diverse bodies and animations. This is largely enabled by our novel application of Transformer-based architectures to this problem: A Transformer in the static network establishes robust draped shapes, and a unique cross-attention design within the dynamic network effectively fuses these static features with motion cues. Unlike methods confined to static poses, our system produces consistent and natural-looking dynamics across a wide range of hair styles and body poses. Furthermore, our method is even capable of producing realistic dynamics for extreme and challenging motion sequences, provided with additional fine-tuning on respective curated training data.

In summary, our contributions are as follows.
    \vspace{-1mm}
\begin{enumerate}
\item A comprehensive suite of physics-informed loss functions designed to maintain hair fidelity, preserve hairstyle characteristics, and promote realistic dynamic behavior.
\item An explicit hair-body penetration resolution mechanism, leveraging an extended IPC-style barrier function to handle negative signed distances and effectively reduce penetrations for full-body meshes.
\item A Transformer-based static network that predicts physics-based, static draped shapes from arbitrary hairstyles, body shapes, and poses, achieving results that generalize to novel hairstyles and body inputs, providing a robust foundation for subsequent dynamic simulation.
\item A dynamic network featuring a novel cross-attention mechanism to fuse static hair geometry with motion cues. This design preserves hairstyle characteristics while accurately predicting detailed and temporally coherent hair motion, and enables efficient fine-tuning for specialized, challenging motion sequences such as sudden head movements.
\end{enumerate}
\section{Related Work}

\subsection{Physics-Based Hair Simulation}
Physics-based hair simulation commonly employs strand-based representations for high fidelity, often utilizing Kirchhoff-based rod models (e.g., Discrete Elastic Rods \cite{bergou2008discrete}, Position-Based Cosserat Rods \cite{kugelstadt2016position}) to capture complex strand mechanics. Significant efforts focus on real-time performance through algorithmic optimizations like ADMM \cite{daviet2023interactive}, efficient collision handling \cite{hsu2024real}, and model enhancements such as the Augmented Mass-Spring (AMS) Model \cite{herrera2024augmented}, which improves traditional mass-spring systems for dense hair. Although high-fidelity, physics-based simulations are computationally expensive and inefficient for applications such as virtual reality that require rapid asset generation, making learning-based approaches more desirable.

\subsection{Learning-Based Hair Simulation}
Learning-based approaches are increasingly applied to hair modeling, generation, animation, and dynamics. For example, NeuWigs \cite{wang2023neuwigs} learns volumetric dynamics from videos, while other methods focus on generating diverse hair geometries. Among these, HAAR \cite{sklyarova2023haar} introduces a text-to-hair method that produces animatable, strand-based 3D models using a latent diffusion process on a geometric texture. Similarly, GroomGen \cite{zhou2023groomgen} presents a generative model for diverse, dense 3D hair geometry by employing a novel hierarchical latent space for both individual strands and overall hairstyles. Other techniques like Neural Haircut \cite{sklyarova2023neural}, TANGLED \cite{long2025tangled}, and PERM \cite{he2024perm} also generate varied geometries using methods ranging from latent textures to parametric models. A significant direction is self-supervised simulation of hair dynamics, where methods like Quaffure \cite{stuyck2024quaffure} use physics-based losses (e.g., from Cosserat rod energy) to achieve real-time, quasi-static results without pre-simulated data, training networks to follow physical principles. However, existing static or quasi-static approaches often fail to capture complex secondary motions and generalize to varied and unseen inputs. Our method addresses these limitations by achieving both realistic dynamics and broad generalizability..

\subsection{Neural Simulation and Advanced Architectures}
Our work builds on the broader context of neural simulation for deformable objects and advancements in network architectures. Inspired by self-supervised methods for cloth drape prediction (e.g. PBNS \cite{bertiche2020pbns}, Neural Cloth Simulation \cite{bertiche2022neural}, SNUG \cite{santesteban2022snug}), which often use body models like SMPL \cite{loper2023smpl} and physics-based losses, our approach extends similar principles to hair. Reinforcing this link, HOOD \cite{grigorev2023hood} proposes a garment-agnostic neural simulator that uses hierarchical graph networks to predict real-time cloth dynamics for a wide range of clothing types. The efficacy of these simulations is enhanced by sophisticated neural network architectures such as GRUs and Transformers, which capture complex spatial-temporal dependencies. For example, HCMT \cite{liu2024hcmt} uses Transformers for collision dynamics. Their combined strengths play a key role in modeling coherent hair behavior and improving generalization. Furthermore, new approaches are emerging, such as using priors from large-scale generative models. For example, PhysDreamer \cite{zhang2024physdreamer} estimates physical properties using video generation priors, offering new simulation approaches.
\section{Method}
\begin{figure*}[ht]
\vspace{-5mm}
    \centering
    \includegraphics[width=\textwidth]{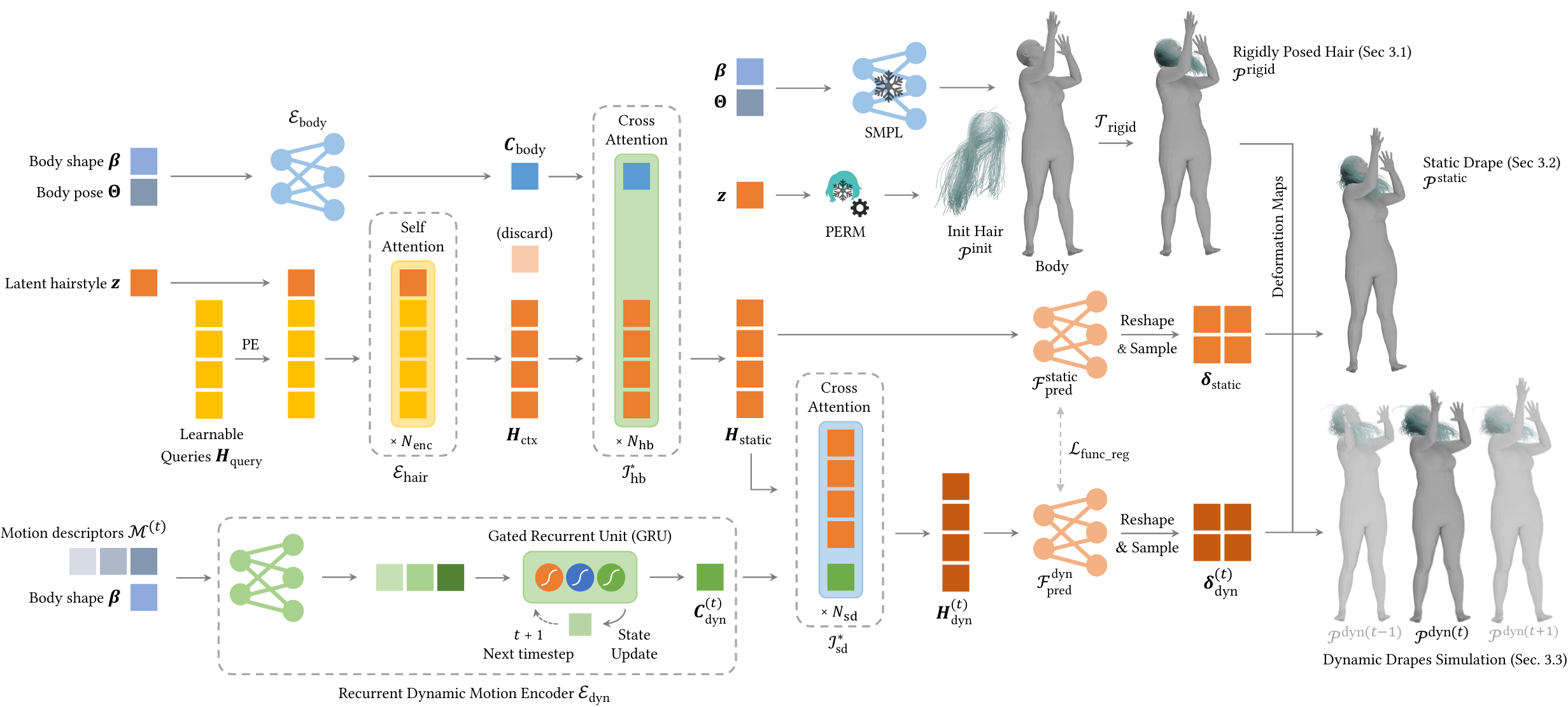}
\vspace{-5mm}
    \caption{\textbf{Method Overview.} Our network architecture features distinct static and dynamic modules. (Top right) Initially, hair strands are generated from hairstyle latent code using a StyleGAN-based model (PERM~\cite{he2024perm}) and then rigidly posed onto the character's body parameterized by SMPL~\cite{loper2023smpl}. (Top left) The static network module processes the hairstyle latent and the body parameters to predict an initial, physically plausible draped shape. (Bottom) Following this, the dynamic network module takes over to simulate expressive hair motion over sequences of body poses. (Right) The initial rigidly posed hairs are shifted by the deformation map predicted by static and dynamic modules to predict the final hair simulation results. Both the static and dynamic networks are trained through a self-supervised approach, guided by physics-based loss functions to ensure realistic and coherent hair animations. }
\vspace{-4mm}
\end{figure*}

Our approach consists of a static network that predicts an initial hair drape and a dynamic network that simulates subsequent motion, both trained with physics-based self-supervision.

\subsection{Representation}

We represent the human body using the Skinned Multi-Person Linear (SMPL) model \cite{loper2023smpl}, parameterized by pose $\bm{\Theta}\in\mathbb{R}^{66}$ (encoding root orientation and 21 joint rotations in axis-angle format) and shape $\bm{\beta}\in\mathbb{R}^{16}$. Hair roots are initially defined over a 2D UV texture map corresponding to the scalp region of the SMPL model's template head.

Our method processes arbitrary synthetic hairstyles using a procedure $\mathcal{G}_\mathrm{hair}$. We represent the full hair geometry as a set of vertices with each vertex denoted as $\bm{p}_{s,v}$, where $s$ indicates the strand index and $v$ represents the vertex index along that strand.\footnote{Throughout this paper, we use superscripts to denote hair state or timestep and use subscripts to denote vertex/strand index in our notation.}
We generate the initial hair $\mathcal{P}^\mathrm{init}$ using a pretrained StyleGAN \cite{he2024perm} conditioned on an input hair latent code $\bm{z} \in \mathbb{R}^{d_z}$ where $d_z=512$. $\mathcal{P}^\mathrm{init}$ is a collection of $M_\mathrm{hair}$ individual hair strands $\{\bm{p}_s^\mathrm{init}\}_{s=1, \dots, M_\mathrm{hair}}$, where each strand $\bm{p}_s^\mathrm{init} \in \mathbb{R}^{N \times 3}$ is defined by $N=100$ ordered vertices in 3D space. Along with the hair strands, $\mathcal{G}_\mathrm{hair}$ also produces their corresponding root UV coordinates, $\bm{uv} \in \mathbb{R}^{M_\mathrm{hair} \times 2}$, effectively mapping the root (first vertex) $\bm{p}_{s,0}^\mathrm{init}$ of each strand $\bm{p}_{s}^\mathrm{init}$ to a unique position on a $M_w \times M_h$ grid (e.g., $32 \times 32$, so $M_\mathrm{hair}=1024$) over the scalp region of a rest body mesh. The entire procedure for generating and preprocessing hair $\mathcal{G}_\mathrm{hair}$ involves StyleGAN sampling, initial strand positioning relative to predefined guide roots on the head at rest pose, and filtering to remove overly short strands. To ensure accurate placement of these generated strands on the specific SMPL head topology, we align the head mesh on which PERM \cite{he2024perm} was trained with our SMPL head template using an Iterative Closest Point (ICP) based approach \cite{besl1992method}. We note that the initial strands \(\mathcal{P}^\mathrm{init}\) are defined relative to a rest pose and mean body shape. For subjects with body shapes that significantly alter the scalp geometry, \(\mathcal{P}^\mathrm{init}\) requires recomputation or projection.

\subsubsection{Rigid Hair Transformation}
To position the generated hair strands $\mathcal{P}^\mathrm{init}$ on a body with an arbitrary shape $\bm{\beta}$ and pose $\bm{\Theta}$, we employ a rigid hair transformation procedure $\mathcal{T}_\mathrm{rigid}$ which produces rigidly posed hair $\mathcal{P}^\mathrm{rigid} = \{\bm{p}_{s,v}^\mathrm{rigid} \}_{s=1}^{M_\mathrm{hair}}$.

We first construct a posed SMPL body mesh and obtain body vertices $\bm{V}_\mathrm{body} \in \mathbb{R}^{N_\mathrm{smpl} \times 3}$ (where $N_\mathrm{smpl}$ is the number of SMPL vertices, typically 6890) and the global affine transformation matrix for the head joint, $\bm{M}_\mathrm{head} \in \mathbb{R}^{4 \times 4}$, relative to the world origin. Each vertex $\bm{p}_{s,v}^\mathrm{init}$ of the generated strands is initially transformed by $\bm{M}_\mathrm{head}$.

To account for non-head joint rotations influencing the posed head shape via Linear Blend Skinning (LBS), we compute a global alignment translation $\Delta\bm{t}_\mathrm{align} \in \mathbb{R}^3$ that is the mean displacement between the vertex positions of the head region in the posed SMPL mesh (a subset of $\bm{V}_\mathrm{body}$) and the vertex positions of the SMPL head from a rest pose $\bm{V}_\mathrm{head}^0$) after transformation by $\bm{M}_\mathrm{head}$. We add $\Delta\bm{t}_\mathrm{align}$ to all transformed hair vertices.

Finally, for precise scalp attachment of the roots, we project the root of each strand \(\bm{p}'_{s,0}\) (after transformation by $\bm{M}_\mathrm{head}$ and $\Delta\bm{t}_\mathrm{align}$) onto the posed body surface $\bm{V}_\mathrm{body}$. We apply the resulting offset $\Delta\bm{t}_{s,0}$ to each vertex of the strand $s$.

The complete transformation for a single vertex $\bm{p}_{s,v}^\mathrm{init}$ to its rigidly posed state $\bm{p}_{s,v}^\mathrm{rigid}$ is thus:
\begin{equation}
\label{eqn:rigid}
\bm{p}_{s,v}^\mathrm{rigid} = \bm{M}_\mathrm{head}\tilde{\bm{p}}_{s,v}^\mathrm{init} + \Delta\bm{t}_\mathrm{align} + \Delta\bm{t}_{s,0}
\end{equation}
where $\tilde{\bm{p}}_{s,v}^\mathrm{init}$ is the homogeneous coordinate counterpart of ${\bm{p}}_{s,v}^\mathrm{init}$. The general function is $\mathcal{P}^\mathrm{rigid} = \mathcal{T}_\mathrm{rigid}(\mathcal{P}^\mathrm{init}, \bm{\beta}, \bm{\Theta})$.

\subsection{Static Hair Transformer Architecture}
\label{sec:static}
We propose a network for predicting hair strand deformations that, when added to rigidly posed hair strands, provides physics-based draped shapes over a single posed body. The network takes the hair latent $\bm{z}$, body shape $\bm{\beta}$, and body pose $\bm{\Theta}$ as inputs, and predicts a deformation map $\bm{D}_\mathrm{static} \in \mathbb{R}^{M_{H_0} \times M_{W_0} \times N \times 3}$ defined on a grid of size $M_{H_0} \times M_{W_0}$ (e.g. $8 \times 8$). The per-vertex deformation $\bm{\delta}_{s,v}^\mathrm{static}$ of strand $s$ is then obtained via a nearest neighbor lookup operation $\mathcal{N}$ on $\bm{D}_\mathrm{static}$, which uses the root UV coordinate of the strand $\mathbf{uv}_s$ to find the corresponding deformation on the coarser $M_{H_0} \times M_{W_0}$ grid. A per-vertex tapering function, $\bm{\mathcal{T}}_\mathrm{taper}(v)$, subsequently scales down the sampled deformation toward the hair root to ensure smoother root attachment and preserve hairstyle characteristics near the scalp.
The final per-vertex static hair position \(\bm{p}_{s,v}^\mathrm{static}\) is then:
\begin{equation}
\begin{aligned}
\bm{p}_{s,v}^\mathrm{static}&=\bm{p}_{s,v}^{\mathrm{rigid}}+\bm{\delta}_{s,v}^\mathrm{static}\ \text{, where}\\
\bm{\delta}_{s,v}^\mathrm{static}&=\mathcal{N}(\bm{D}_\mathrm{static}(\bm{z},\bm{\beta},\bm{\Theta}), \mathbf{uv}_s)\odot \bm{\mathcal{T}}_\mathrm{taper}(v).
\end{aligned}
\label{eqn:staticnetwork}
\end{equation}

Our Transformer-based network first encodes hair style and body information separately, then fuses these representations using cross-attention to predict the deformation map $\bm{D}_\mathrm{static}$. Formally, the static network comprises the following modules: Hair encoder $\mathcal{E}_{\mathrm{hair}}$, body encoder $\mathcal{E}_{\mathrm{body}}$, hair-body cross-attention blocks $\mathcal{I}_{hb}^*$, and prediction head $\mathcal{F}_\mathrm{pred}^{\mathrm{static}}$.
\begin{equation}
    \begin{aligned}
        \bm{H}_{\mathrm{ctx}} &= \mathcal{E}_{\mathrm{hair}}(\bm{z}) \\
        \bm{C}_{\mathrm{body}} &= \mathcal{E}_{\mathrm{body}}([\bm{\beta},\bm{\Theta}]) \\
        \bm{H}_{\mathrm{static}} &= \mathcal{I}_{hb}^*(\bm{H}_{\mathrm{ctx}},\bm{C}_{\mathrm{body}}) \\
        \bm{D}_{\mathrm{static}} &= \mathrm{Reshape}(\mathcal{F}_\mathrm{pred}^{\mathrm{static}}(\bm{H}_{\mathrm{static}}))
    \end{aligned}
\label{eqn:static_components_brief}
\end{equation}

\subsubsection{Hair Encoder} 
\label{sec:hair-encoder}
Given a hair latent $\bm{z} \in \mathbb{R}^{d_z}$, the hair encoder $\mathcal{E}_{\mathrm{hair}}$ generates $L = M_{H_0} \times M_{W_0}$ (e.g., $8 \times 8 = 64$) hair context features $\bm{H}_{\mathrm{ctx}} \in \mathbb{R}^{L \times d_m}$. We set the internal feature dimension of our Transformer modules $d_m=d_z=512$, the same size as the hair latent. These $L$ features correspond to the latent information for each cell in the $M_{H_0} \times M_{W_0}$ grid that underpins the deformation map $\bm{D}_\mathrm{static}$. The hair encoder includes the following components. \vspace{1ex}

\noindent\emph{Learnable Query Vectors:} $L$ learnable query vectors $\bm{H}_\mathrm{query}\in\mathbb{R}^{L\times d_m}$ are augmented with 2D positional encoding (PE) to provide initial spatial awareness for each region on the \(M_{H_0}\times M_{W_0}\) grid. \vspace{1ex}

\noindent\emph{Self-Attention Layers:} The queries $\bm{H}_\mathrm{query}$ are concatenated with the hair latent $\bm{z}$ to form an input token sequence of length $(L+1)$ to $N_\mathrm{enc}$ self-attention layers, forming a Transformer encoder that exchanges information between the query vectors and the hair latent. \vspace{1ex}

\noindent\emph{Output:} The resulting $(L+1)$ token sequence is processed by discarding the token corresponding to $\bm{z}$; the remaining $L$ tokens form the hair context features $\bm{H}_{\mathrm{ctx}}$.
\vspace{-1ex}
\subsubsection{Body Encoder}
\label{sec:body-encoder}
A multilayer perceptron (MLP) $\mathcal{E}_{\mathrm{body}}$ projects the concatenated body shape $\bm{\beta}$ and pose $\bm{\Theta}$ vectors into a global body context vector $\bm{C}_\mathrm{body}\in\mathbb{R}^{1\times d_m}$.
\vspace{-1ex}

\subsubsection{Hair-Body Cross-Attention blocks}
\label{sec:cross-attention}
To predict hair deformations in response to to body characteristics, a series of $N_{hb}$ cross-attention blocks, collectively denoted as $\mathcal{I}_{hb}^*$, modulate the hair context features $\bm{H}_{\mathrm{ctx}}$ using the body context $\bm{C}_\mathrm{body}$. Each block $l \in [1, N_{hb}]$ employs a multi-head cross-attention layer where $\bm{H}_{\mathrm{ctx}}^{(l-1)}$ (hair features from the previous block, or $\bm{H}_{\mathrm{ctx}}^{(0)}=\bm{H}_{\mathrm{ctx}}$) acts as queries, and $\bm{C}_\mathrm{body}$ serves as keys and values, allowing hair features to selectively integrate global body information. Each cross-attention layer is followed by a position-wise feed-forward network (FFN). The output of the final block is $\bm{H}_{\mathrm{static}} \in \mathbb{R}^{L \times d_m}$.

\subsubsection{Prediction Head}
\label{sec:staticprediction}
A prediction head $\mathcal{F}_\mathrm{pred}^{\mathrm{static}}$ composed of MLPs transforms the body-aware hair tokens $\bm{H}_\mathrm{static}\in\mathbb{R}^{L\times d_m}$ from cross-attention blocks into a flat list of deformation vectors (shaped $L\times(N\times3)$), which is then structured via reshaping into the final deformation map $\bm{D}_{\mathrm{static}}\in\mathbb{R}^{M_{H_0} \times M_{W_0} \times N \times 3}$.


Leveraging physics-based loss functions (detailed in \cref{sec:loss}), our static Transformer architecture accurately predicts physically plausible, quasi-static hair drapes for diverse hairstyles, body poses and shapes. It robustly resolves penetrations and preserves hair properties, providing a strong basis for dynamic motion.

\subsection{Dynamic Hair Network Architecture}
\label{sec:dynamic}
Inspired by Neural Cloth Simulation \cite{bertiche2022neural}, we design a dynamic network that integrates temporal information over a sequence of body poses with the hair features per pose predicted by the static network in \cref{sec:static} to simulate hair motion.\vspace{1ex}

\subsubsection{Recurrent Dynamic Motion Encoder} To capture temporal dynamics, our network processes sequential motion descriptors encoding both current and historical body poses. For timestep $t$, we extract a body pose sequence within a time window of width $T_\mathrm{w}$, denoted as  $\bm{\vartheta}^{(t)}=\{\bm{\Theta}^{(t-T_\mathrm{w}+1)},\dots,\bm{\Theta}^{(t)}\}$. We then compute the motion descriptors $\mathcal{M}^{(t)}=\{\bm{m}^{(t-T_{\mathrm{w}}+1)},\dots,\bm{m}^{(t)}\}=\mathcal{F}_\mathrm{motion}(\bm{\vartheta^{(t)}},\bm{\beta},\Delta t)$, which captures first-order temporal derivatives of static pose characteristics (joint rotations, unposed gravity) and second-order local joint accelerations, ensuring $\mathcal{M}^{(t)}$ retains the length of $\bm{\vartheta}^{(t)}$. The motion descriptors $\mathcal{M}^{(t)}$, along with the body shape $\bm{\beta}$, are processed through fully-connected layers and a subsequent gated recurrent unit (GRU). The GRU integrates the previous hidden state $\bm{C}_\mathrm{dyn}^{(t-1)}$ to produce a latent code $\bm{C}_\mathrm{dyn}^{(t)}=\mathcal{D}_\mathrm{dyn}(\bm{C}_\mathrm{dyn}^{(t-1)},\mathcal{M}^{(t)})$ that encapsulates dynamic information of the current timestep while serving as the hidden state for the next timestep.

\subsubsection{Static-Dynamic cross-attention blocks} Given the static hair features $\bm{H}_\mathrm{static}^{(t)}$, our static-dynamic cross-attention blocks $\mathcal{I}_{sd}^*$ neural-simulate hair dynamics by cross-attending $\bm{H}_\mathrm{static}^{(t)}$ to the dynamic latent code $\bm{C}_\mathrm{dyn}^{(t)}$ and predicting the dynamic hair features $\bm{H}_\mathrm{dyn}^{(t)}$. Specifically, similar to \cref{sec:cross-attention}, we employ $N_{sd}$ cross-attention and FFN layers, where the hair features $\bm{H}_\mathrm{static}^{(t)}$ act as queries and the dynamic latent $\bm{C}_\mathrm{dyn}^{(t)}$ serves as keys and values, effectively interacting hair features with the dynamic context.

\subsubsection{Prediction Head and Reshape} Similar to \cref{sec:staticprediction}, the dynamic hair features $\bm{H}_\mathrm{dyn}^{(t)}$ are processed by a final MLP $\mathcal{F}_\mathrm{pred}^{\mathrm{dyn}}$ and then reshaped into the final deformation map $\bm{D}_{\mathrm{dyn}}\in\mathbb{R}^{M_{H_0} \times M_{W_0} \times N \times 3}$.

\subsubsection{Functional Regularization}
\label{sec:funcreg}
Our dynamic network aims to generate dynamic hair features $\bm{H}_\mathrm{dyn}^{(t)}$ by augmenting static hair features $\bm{H}_\mathrm{static}^{(t)}$ with dynamic context $\bm{C}_\mathrm{dyn}^{(t)}$ via cross-attention layers. Since $\bm{H}_\mathrm{static}^{(t)}$ already encapsulates rich latent hair properties from the pretrained static network, we seek to encourage $\bm{H}_\mathrm{dyn}^{(t)}$ to evolve within a compatible latent space. To this end, we employ a functional regularization strategy~\cite{hinton2015distilling, garg2020functional}, which maintains consistency between the original static prediction head $\mathcal{F}_\mathrm{pred}^{\mathrm{static}}$ and the dynamic prediction head $\mathcal{F}_\mathrm{pred}^{\mathrm{dyn}}$. Specifically, we enforce that its functional behavior on the original static features $\bm{H}_\mathrm{static}^{(t)}$ remains similar to that of the (frozen) static head:
\begin{equation}
    \mathcal{L}_\mathrm{func\_reg}^{(t)} = \lambda_\mathrm{fr} \cdot \mathrm{MSE}\left(\mathcal{F}_\mathrm{pred}^{\mathrm{dyn}}(\bm{H}_\mathrm{static}^{(t)}), \mathcal{F}_\mathrm{pred}^{\mathrm{dyn}}(\bm{H}_\mathrm{static}^{(t)})\right)
\label{eqn:funcreg_loss_t}
\end{equation}
This implicitly guides the trainable layers producing $\bm{H}_\mathrm{dyn}^{(t)}$ to generate features that are representationally consistent with $\bm{H}_\mathrm{static}^{(t)}$, encouraging the preservation of established static feature characteristics within the dynamic representations without drastic deviations.

\subsection{Physics-Based Loss for Hair Dynamics}
\label{sec:loss}
Inspired by the self-supervision principle of cloth drape simulation such as PBNS~\cite{bertiche2020pbns}, we train our static and dynamic networks by learning to minimize the total potential energy of hair strands on a single body or over a sequence of posed bodies. We carefully define each potential energy component and constraint as loss functions for training to ensure the physical plausibility of our simulation results in alignment with the behaviors of real-world hair strands: They are nearly inextensible, retain their natural shapes under gravity (sag-free), and do not penetrate into the body. Let $\mathcal{P}^{(t)} = \{\bm{p}_{s,v}^{(t)} | s=1, \dots, M, v=1, \dots, N\}$ be the set of all hair vertex positions for a hairstyle of $M$ strands, each with $N$ vertices, at timestep $t$. Our loss functions collectively define the total energy of the system for the hair strands.

The total loss for the static network evaluated at timestep $t$ is:
\begin{equation}
\mathcal{L}_\mathrm{tot}^{\mathrm{static}(t)} = \mathcal{L}_{s}^{(t)} + \mathcal{L}_{b}^{(t)} + \mathcal{L}_\mathrm{aux}^{(t)} + \mathcal{L}_\mathrm{smooth}^{(t)} + \mathcal{L}_\mathrm{gravity}^{(t)} + \mathcal{L}_\mathrm{hb}^{(t)} + \mathcal{L}_\mathrm{hh}^{(t)} + \mathcal{L}_\mathrm{root}^{(t)}
\label{eqn:total-energy-static}
\end{equation}
For the dynamic network, the total loss at timestep $t$ additionally includes an inertia term and a functional regularization term:
\begin{equation}
\mathcal{L}_\mathrm{tot}^{\mathrm{dyn}(t)} = \mathcal{L}_\mathrm{tot}^{\mathrm{static}(t)} + \mathcal{L}_\mathrm{inertia}^{(t)} + \mathcal{L}_\mathrm{func\_reg}^{(t)}
\label{eqn:total-energy-dynamic}
\end{equation}
When training the dynamic network on a sequence of poses, the overall loss for the sequence is typically the sum or average of $\mathcal{L}_\mathrm{tot}^{\mathrm{dyn}(t)}$ computed over all timesteps in the sequence for which all terms (especially $\mathcal{L}_\mathrm{inertia}^{(t)}$) are well-defined. Each loss component, detailed in the following subsections, is evaluated based on the relevant hair vertex positions (e.g., $\mathcal{P}^{(t)}$, $\mathcal{P}^{(t-1)}$, $\mathcal{P}^{(t-2)}$), rigidly posed hair $\mathcal{P}^{\mathrm{rigid}(t)}$, and network features like $\bm{H}_\mathrm{static}^{(t)}$.

We fix the hair root and its adjacent vertex for each strand, i.e. $\bm{p}_{s,0}^{(t)}$ and $\bm{p}_{s,1}^{(t)}$, when predicting the deformation relative to $\mathcal{P}^{\mathrm{rigid}(t)}$, and exclude them from gradient computation.
We also exclude these fixed vertices from the gravity and inertia loss functions, acting only on the masses of the $N^\mathrm{mov}=N-2$ \emph{moving} vertices, while preserving them in the computation of inextensibility and bending losses.
We use $v\in[0,N-1]$ to index the hair vertices in the \textit{full} strand and $v'\in[0,N^\mathrm{mov}-1]$ to index the moving vertices (corresponding to $v \in [2,N-1]$). 
Since we may predict deformations for guide strands representing bundles of finer hairs, physical parameters like mass can be scaled to reflect aggregated properties.

\subsubsection{Inextensibility Loss}
To penalize changes in hair segment lengths from their rest configuration $\mathcal{P}^\mathrm{init}$, 
we use a spring-like model and define an inextensibility loss resembling the elastic potential energy:
\vspace{-0.75ex}
\begin{equation}
\begin{aligned}
\mathcal{L}_{\mathrm{s}}^{(t)} &= k_{\mathrm{s}} \frac{1}{M(N-1)} \sum_{s=1}^{M} \sum_{i=0}^{N-2} (\|\bm{p}_{s,i+1}^{(t)} - \bm{p}_{s,i}^{(t)}\| - \overline{l}^0_{s,i})^2
\end{aligned}
\label{eqn:inextensibility_loss_t}
\end{equation}
\vspace{-0.75ex}
where
$\overline{l}^0_{s,i} = \|\bm{p}_{s,i+1}^{0} - \bm{p}_{s,i}^{0}\|$ is the rest length of segment $i$ of strand $s$ in the initial generated hair.

\subsubsection{Hair Shape Losses}
We use three loss functions to penalize drastic changes in hair shapes, encouraging strands to behave naturally under gravity while attempting to maintain their characteristic curvature. We omit twist computation as guide strands primarily capture global hairstyle information.

Our \emph{bending loss} accumulates the turning angles between consecutive hair segments to penalize sharp turns:
\begin{align}
\mathcal{L}_{\mathrm{b}}^{(t)} &= k_{\mathrm{b}} \sum_{s=1}^{M} \sum_{i=0}^{N-3} \alpha_{s,i}^{(t)}\ \text{, where} \label{eqn:bending_loss_t}\\
\alpha_{s,i}^{(t)} &= \begin{cases} \arccos(\mathrm{clamp}(\theta_{s,i}^{(t)}, -1+\epsilon, 1-\epsilon)) & \mathrm{if}\ C_{\mathrm{valid}}(s,i,t) \\ 0 & \mathrm{otherwise} \end{cases}\notag\\
\bm{e}_{s,i}^{(t)} &= \bm{p}_{s,i+1}^{(t)} - \bm{p}_{s,i}^{(t)} \notag\\
\theta_{s,i}^{(t)} &= \frac{\bm{e}_{s,i}^{(t)}}{\|\bm{e}_{s,i}^{(t)}\|+\epsilon} \cdot \frac{\bm{e}_{s,i+1}^{(t)}}{\|\bm{e}_{s,i+1}^{(t)}\|+\epsilon} \notag\\
C_{\mathrm{valid}}(s,i,t) &\equiv (\|\bm{e}_{s,i}^{(t)}\| > \epsilon \mathrm{\ and\ } \|\bm{e}_{s,i+1}^{(t)}\| > \epsilon) \notag
\end{align}

To counteract the tendency of hair to become overly straight and to preserve characteristics of the input hairstyle, we use an \emph{auxiliary loss} to penalize significant deviations from the rigidly posed hair configuration $\mathcal{P}^{\mathrm{rigid}(t)}$:
\begin{equation}
\mathcal{L}_{\mathrm{aux}}^{(t)} = \lambda_{\mathrm{aux}} \frac{1}{3MN}\sum_{s=1}^{M} \sum_{v=0}^{N-1} \|\bm{p}_{s,v}^{(t)} - \bm{p}^{\mathrm{rigid}(t)}_{s,v}\|^2_2
\label{eqn:aux_loss_t}
\end{equation}

A \emph{smoothness loss} is applied to penalize jitter and approximate curvature by minimizing the difference between adjacent edges:
\begin{equation}
\mathcal{L}_{\mathrm{smooth}}^{(t)} = \lambda_{\mathrm{smooth}} \frac{1}{M(N-2)} \sum_{s=1}^{M} \sum_{v=1}^{N-2} \| \bm{p}_{s,v+1}^{(t)} - 2\bm{p}_{s,v}^{(t)} + \bm{p}_{s,v-1}^{(t)} \|^2
\label{eqn:smoothness_loss_t}
\end{equation}

\subsubsection{Gravity Loss}
This loss applies a downward force by considering the gravitational potential energy of the moving vertices:
\begin{equation}
\mathcal{L}_{\mathrm{gravity}}^{(t)} = - \sum_{s=1}^{M} \sum_{v'=0}^{N_{\mathrm{mov}}-1} (m_{s,v'} \bm{g}) \cdot \bm{p}^{\mathrm{mov}(t)}_{s,v'}
\label{eqn:gravity_loss_t}
\end{equation}
where $\bm{p}^{\mathrm{mov}(t)}_{s,v'}$ are the positions of the moving vertices of strand $s$ at time $t$, and $m_{s,v'}$ is the mass of moving vertex $v'$.

\subsubsection{Contact Losses}
\label{sec:contact}
We use the same formulation for $\mathcal{L}_\mathrm{hb}^{(t)}$ (hair-body contact) and $\mathcal{L}_\mathrm{hh}^{(t)}$ (hair-hair contact), where $d^{(t)}$ is the signed distance to the body surface or the distance between hair segments, respectively. Inspired by the barrier function in codimensional incremental potential contact (C-IPC) \cite{li2020codimensional}, we define a piecewise barrier function $B(d)$ that effectively handles penetrating hair segments, by extrapolating the logarithmic barrier into negative signed distances (penetration) using a quartic function, while ensuring $\mathcal{C}^2$-continuity from a join point $d_b=\xi+b_p\hat d$ where $b_p\in(0,1)$. The hard barrier ($\xi$) is set at the hair diameter for hair-hair contacts and 1mm for hair-body contacts, and the soft barrier ($\hat{d}$) is set to 1.5 times the respective hard barrier distance. 
\begin{equation}
\begin{aligned}
\mathcal{L}_\mathrm{contact}^{(t)} &=\lambda_\mathrm{contact}B(d^{(t)})\ \text{, where } B(d^{(t)})=
\begin{cases}
  B_0(d^{(t)}) & \mathrm{if\ } d^{(t)} \ge d_b \\
  B_Q(d^{(t)}) & \mathrm{if\ } d^{(t)} < d_b
\end{cases}\\
B_0(d^{(t)})&=\max(0,-( (d^{(t)})^2-(\xi+\hat d)^2)^2 \log{\frac{(d^{(t)})^2-\xi^2}{2\xi\hat d+\hat d^2}})\\
B_Q(d^{(t)})&=\max(0,f_b+f_b'\tau+D_f \tau^2+a_3\tau^3+a_4\tau^4), \text{ with } \tau=d^{(t)}-d_b
\end{aligned}
\label{eqn:contact_loss_t}
\end{equation}
The coefficients of the quartic function are derived from $B_0(d)$ and its derivatives at $d_b$ to ensure $\mathcal{C}^2$-continuity. 
\begin{equation}
    f_b=B_0(d_b),f_b'=\left.{\frac{\partial B_0}{\partial d'}}\right|_{d'=d_b}, f_b''=\left.{\frac{\partial^2 B_0}{\partial d'^2}}\right|_{d'=d_b}
\end{equation}
The two remaining DoFs $a_3$ and $a_4$ are resolved by defining the function's behavior at a predetermined deep penetration distance $t_0$, which guarantees a smoothly increasing penalty as penetration depth increases.
\begin{equation}
\begin{aligned}
&D_f = \frac{f_b''}{2}, &t_0=-(\xi+\hat d)-d_b\\
&E_1 = -(f_b + f'_b t_0 + D_f t_0^2), &E_2 = -(f'_b + 2 D_f t_0)\\
&a_3 = \frac{4 E_1}{t_0^3} - \frac{E_2}{t_0^2}, &a_4 = \frac{E_2}{t_0^3} - \frac{3 E_1}{t_0^4}
\end{aligned}
\end{equation}

\begin{figure}[H]
\centerline{\includegraphics[width=\columnwidth]{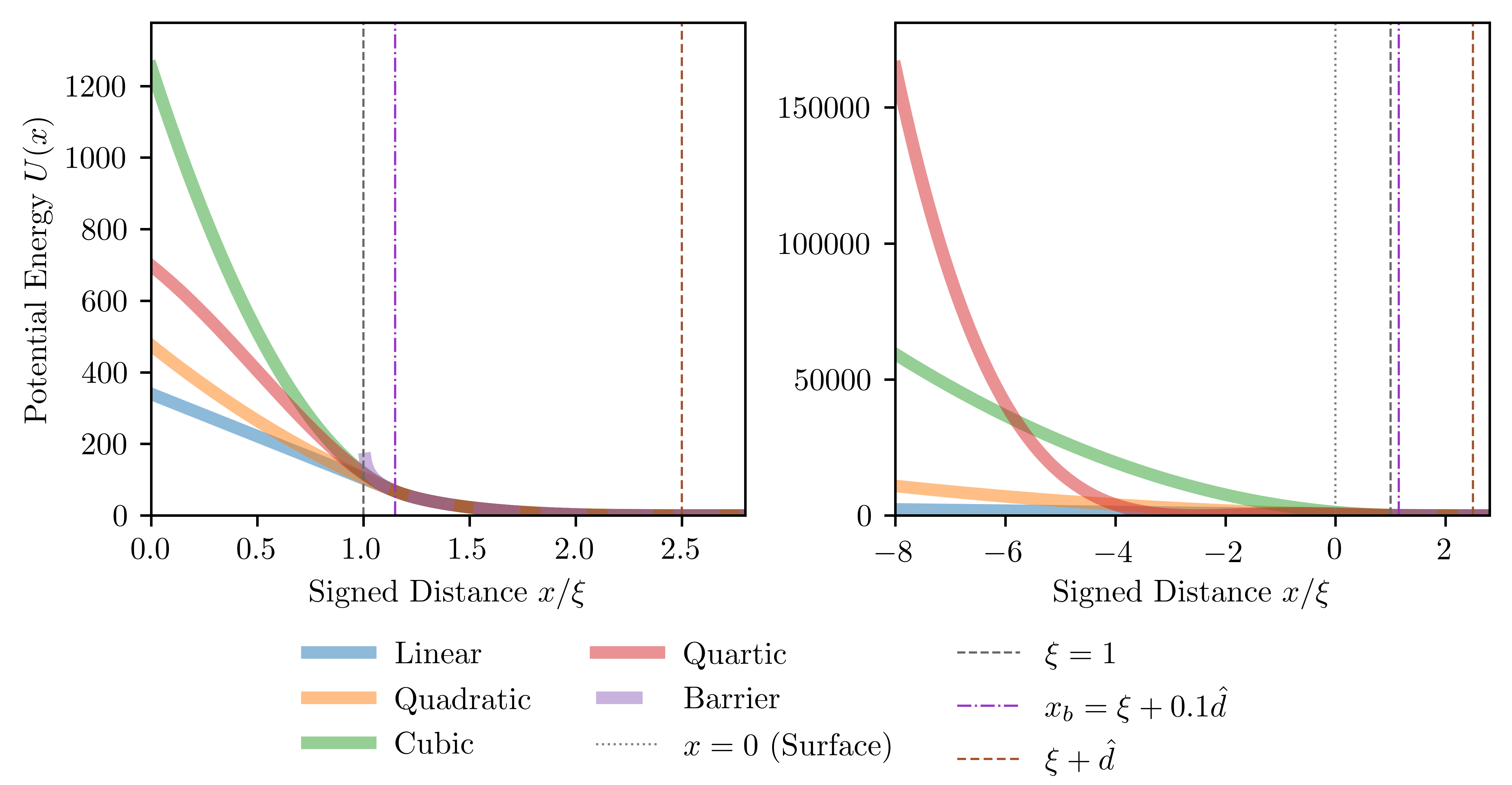}}
\caption{Extrapolated barrier function of various orders.}
\label{fig:barrier}
\end{figure}

Figure \ref{fig:barrier} demonstrates the effect of different polynomial orders on the behavior of the barrier functions used to resolve penetration. Although a cubic function can exhibit a large penalty for relatively small penetrations, its performance for deep penetrations is highly dependent on the careful selection of a free parameter. Despite experimenting with a range of cubic coefficients and selecting the one that performs best, our quartic function consistently demonstrates a greater penalty in deep penetrations compared to the cubic function, especially beyond the predetermined deep penetration distance $t_0$ which essentially mirrors the joining point in the negative sign distance.

We further apply a \emph{root alignment loss} to penalize penetration near the hair root by encouraging alignment with scalp normals:
\begin{align}
    \mathcal{L}_{\mathrm{root}}^{(t)} &= \lambda_{\mathrm{root}} \left(1 - \frac{1}{M k_{\mathrm{align}}'} \sum_{s=1}^{M} \sum_{i=1}^{k_{\mathrm{align}}'} \left( \hat{\bm{d}}_{s,i}^{(t)} \cdot \bm{n}^{\mathrm{scalp}(t)}_{s} \right) \right)\ \text{, where} 
\label{eqn:root_loss_t}\\[-1ex]
    \hat{\bm{d}}_{s,i}^{(t)}&=\frac{\bm{p}_{s,i+1}^{(t)}-\bm{p}_{s,i}^{(t)}}{\|\bm{p}_{s,i+1}^{(t)}-\bm{p}_{s,i}^{(t)}\|+\epsilon_\mathrm{norm}} \notag
\end{align}
and $k_{\mathrm{align}}'$ is the number of initial moving segments to consider.

\subsubsection{Inertia Loss}
We define the inertia loss following \cite{bertiche2022neural} where the current state $x_t$ is predicted based on previous states $x_{t-1}, x_{t-2}$. Let $\bm{p}^{\mathrm{mov, proj}(t)}_{s,v'}$ be the projected position of the moving vertex $v'$ of strand $s$ at time $t$, defined as $\bm{p}^{\mathrm{mov, proj}(t)}_{s,v'} = 2\bm{p}^{\mathrm{mov}(t-1)}_{s,v'} - \bm{p}^{\mathrm{mov}(t-2)}_{s,v'}$. The loss is:
\begin{equation}
\mathcal{L}_{\mathrm{inertia}}^{(t)} = \frac{\lambda_\mathrm{inertia}}{M N_{\mathrm{mov}}} \sum_{s=1}^{M} \sum_{v'=0}^{N_{\mathrm{mov}}-1} \frac{m_{s,v'}}{2(\Delta t)^2} \|\bm{p}^{\mathrm{mov}(t)}_{s,v'} - \bm{p}^{\mathrm{mov, proj}(t)}_{s,v'}\|^2
\label{eqn:inertia_loss_t}
\end{equation}
It is important that gradients are not back-propagated through $\bm{p}^{\mathrm{mov}(t-1)}$ and $\bm{p}^{\mathrm{mov}(t-2)}$ when computing this loss. Freezing gradients for previous hair states ensures that the model learns true dynamic responses to motion, preventing backward information flow and artificial smoothing of past movement, particularly with rapid head oscillations.

\section{Results}
\subsection{Implementation Details}
All models are implemented in PyTorch, with training and inference performed on a single NVIDIA RTX A6000 GPU. Both our static and dynamic networks are trained using the Adam optimizer with a learning rate of $10^{-4}$ and a batch size of 32. We use the Dance Motion Capture Dataset from AMASS \cite{AMASS:2019}, which comprises more than 50,000 distinct poses for each of the 20 dancers with varied body shapes.

\subsection{Training Details}
We train our static and dynamic networks separately in two stages.

\subsubsection{Training Static Network}
To enhance generalization across diverse hair latents, body poses, and shapes, we employ a four-phase curriculum learning strategy to train the static network. Per-vertex deformations $\bm{\delta}_{s,v}^{\mathrm{static}}$ are randomly initialized and clipped at $0.01$.\vspace{1ex}

\noindent\emph{Phase 0:} Training begins with a single posed body at rest, using a mean body shape from a fixed representative set of $\bm{\beta}$'s and a single randomly sampled hair latent. For the contact penalty (\cref{sec:contact}), we start at $b_p=0.9$ and gradually decrease to $b_p=0.01$ throughout training to eliminate hair penetrations into the body mesh.\vspace{1ex}

\noindent\emph{Phase 1:} Starting from the rest pose, the body shape, and the hair latent from Phase 0, the head rotations gradually expand along three dimensions via sampling along ellipsoids with increasing radii.\vspace{1ex}

\noindent\emph{Phase 2:} The training set expands to include 1,000 hair latents and full-body movements from a single dancer.\vspace{1ex}

\noindent\emph{Phase 3:} This final phase uses 20,000 hair latents, pose sequences from all dancers, and augments body shapes by sampling $\bm{\beta}$ from the entire set of dancers and adding Gaussian noise.\vspace{1ex} 

\subsubsection{Training Dynamic Network}
In this stage, we freeze the pretrained static network and initialize the dynamic prediction head with the parameters of the static prediction head.
We use $\Delta t=1/60$ seconds and a GRU window size of $T_w=10$. To ensure sufficient motion from pose sequences, we enhance dynamics by sampling one frame every 2-5 frames, maintaining a consistent interval per sequence. Additionally, we filter out insufficiently dynamic pose sequences by computing a "dynamic score" based on joint movements from 500 randomly sampled sequences and applying the median score as a filtering threshold. 

\subsubsection{Tapering}
\label{sec:taper}
For the raw deformation outputs of the static and dynamic networks, we adopt a tapering operation to scale down the per-vertex deformation according to the vertex position relative to the hair root. The tapering function encourages hair vertices nearer the root to have less deformation, which ensures smoother root attachment and preserves hairstyle characteristics near the scalp. Specifically, for a vertex with indices $v$ ranging from $1$ (closer to the root) to $N$ on a hair strand, its deformation is scaled by a constant factor linearly spaced from $0$ to $1$, i.e.
\[
\bm{D}_{s,v} = \frac{v-1}{N-1}\hat{\bm{D}}_{s,v},
\]
where $s,v$ are strand and vertex indices, respectively, and $\hat{\bm{D}}_{s,v}$ denotes the raw deformation output of the static/dynamic network.

\subsection{Quantitative Comparison}
We select the following as our baselines: (1) A re-implementation of Quaffure adapted for our hair latent input format, and (2) direct optimization of hair positions using either Adam or L-BFGS.
Here, we elaborate on the implementation details of the baseline methods that we compare our method with in Table 1 of our main paper.

\subsubsection{Quaffure (Adapted)}
A direct comparison with the original Quaffure model~\cite{stuyck2024quaffure} is not feasible because its deformation network was trained on a latent space generated by its own autoencoder. This latent space is fundamentally incompatible with the PERM~\cite{he2024perm} StyleGAN latent code used in our pipeline.

To create a fair baseline, we therefore re-implemented their method in PyTorch and adapted their Groom Deformation Decoder to accept our latent codes. This involved replacing their native latent input with ours and re-training the network. The architecture of our adapted decoder is as follows:
\begin{enumerate}
    \item \textit{Input Processing:} We concatenate our PERM hair latent $\bm{z}$ with the body shape $\bm{\beta}$ and pose $\bm{\Theta}$ parameters. A linear layer then processes this combined vector, reshaping it into an initial $512\times4\times4$ feature map.
    
    \item \textit{Upsampling:} The map is then upsampled to $64\times32\times32$ through three transposed convolutional layers.
    
    \item \textit{Output Layer:} A final convolutional layer produces the deformation map, shaped $(N\times3)\times32\times32$, to which we apply the tapering operation from Section~\ref{sec:taper}.
\end{enumerate}

\subsubsection{Direct Optimization} This baseline directly optimizes all hair vertex positions from our physics-based losses using optimizers including Adam and L-BFGS. In particular, we use the rigidly posed hair strands $\mathcal{P}^\mathrm{rigid}$ as the initial positions, and consider all hair vertices except the first two on each strand as optimizable parameters. In each iteration, we calculate physics-based loss functions the same as the training loss of our network, and use the gradients to optimize the hair vertex positions using Adam or L-BFGS. The optimization iterations stop when the loss doesn't decrease for more than a specific number (we chose 200) of iterations.

We use the test scene from Figure \ref{fig:cnn} (middle row) featuring long hair on a tilted head, causing all hair tips to be initially penetrating. This demanding scenario requires robust penetration resolution to achieve a natural drape. As shown in \cref{tab:benchmark}, HairFormer significantly outperforms baselines in both static and dynamic performance, achieving inference times of mere milliseconds. In particular, the dynamic network excels at resolving penetrations and preserving hair length. This superior performance stems from its training on a richer context of pose sequences, which provides more information for optimal drape prediction within a sequence. In contrast, the adapted Quaffure demonstrates a complete inability to learn to decode latent hair, introducing even more hair penetrations than in the initial state. This highlights that our network design learns a general hair representation alongside the physics needed for penetration resolution and physically plausible drapes, allowing them to overcome local minima where simple optimizers get stuck. 
\begin{table}[htbp] 
\vspace{-2ex}
\centering
\footnotesize
\begin{tabular}{@{}lccc@{}}
\toprule
Method & \begin{tabular}[c]{@{}c@{}}Time \\ (ms/drape)$\downarrow$ \end{tabular} & \begin{tabular}[c]{@{}c@{}}Penetration \\ (\%)$\downarrow$ \end{tabular} & \begin{tabular}[c]{@{}c@{}}Length Change \\ (\%)$\downarrow$ \end{tabular} \\
\midrule
Initial State & - & $15.448$ & - \\
HairFormer (Static) & $4.627$ & $11.461$ & $6.791$ \\
HairFormer (Dynamic) & $12.197$ & $6.751$& $1.752$ \\
Quaffure (Adapted) & $27.185$ & $17.022$ & $8.259$ \\
\midrule
Adam & $5194.58$ & $14.749$ & $0.774$ \\
L-BFGS & $2208.34$ & $14.619$ & $0.677$\\
\bottomrule
\end{tabular}
\caption{Key Metrics Comparison. }
\label{tab:benchmark}
\end{table}
\vspace{-10mm}

\subsection{Qualitative Results}
\begin{figure}[H]
\vspace{-10mm}
\centerline{\includegraphics[width=\columnwidth]{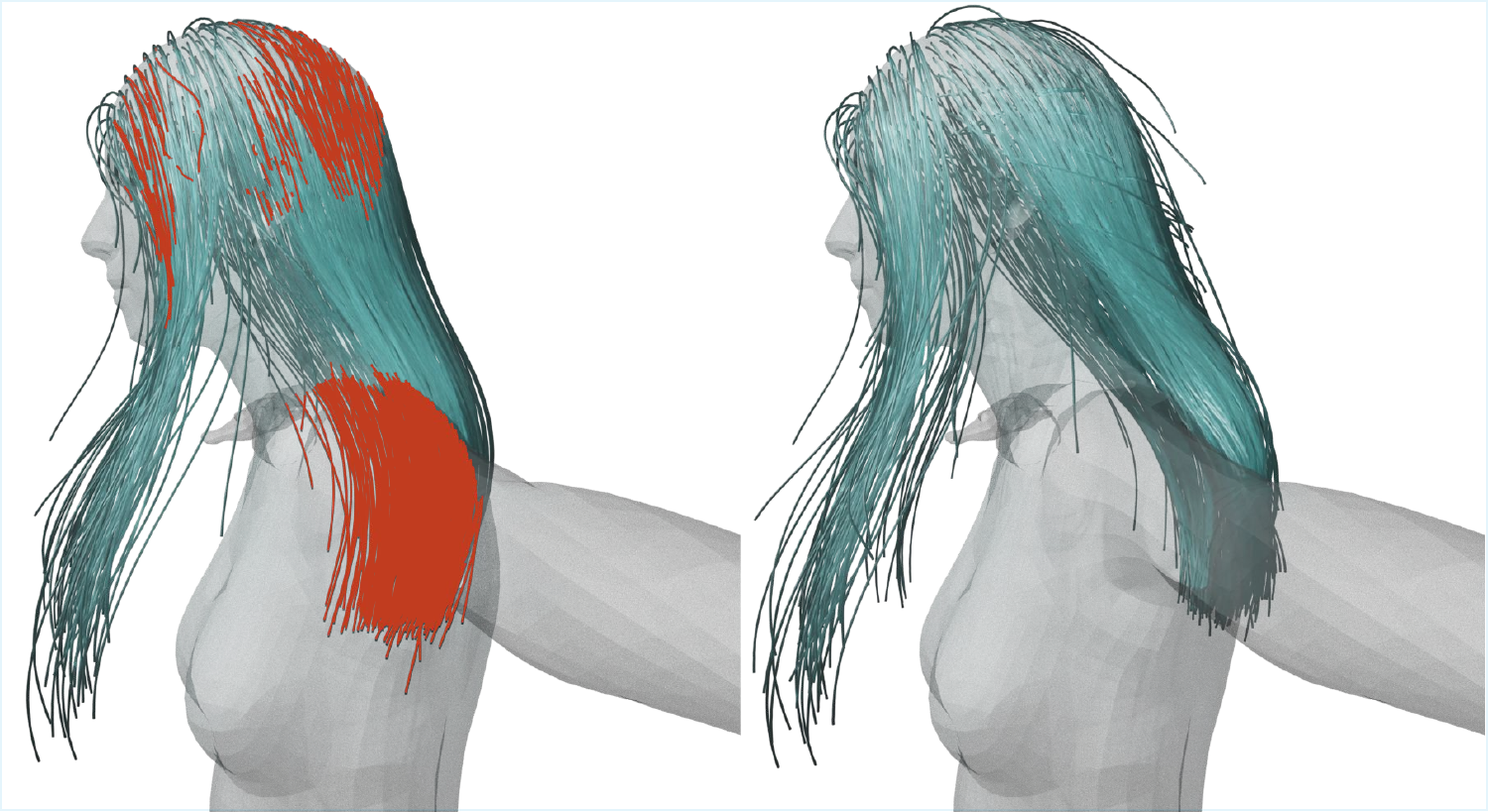}}
\vspace{-1.5ex}\caption{Rigidly-posed hair versus simulated hair in Phase 0 training of the static network.}
\label{fig:penetration}
\end{figure}

\begin{figure*}
\centering
\includegraphics[width=0.84\textwidth]{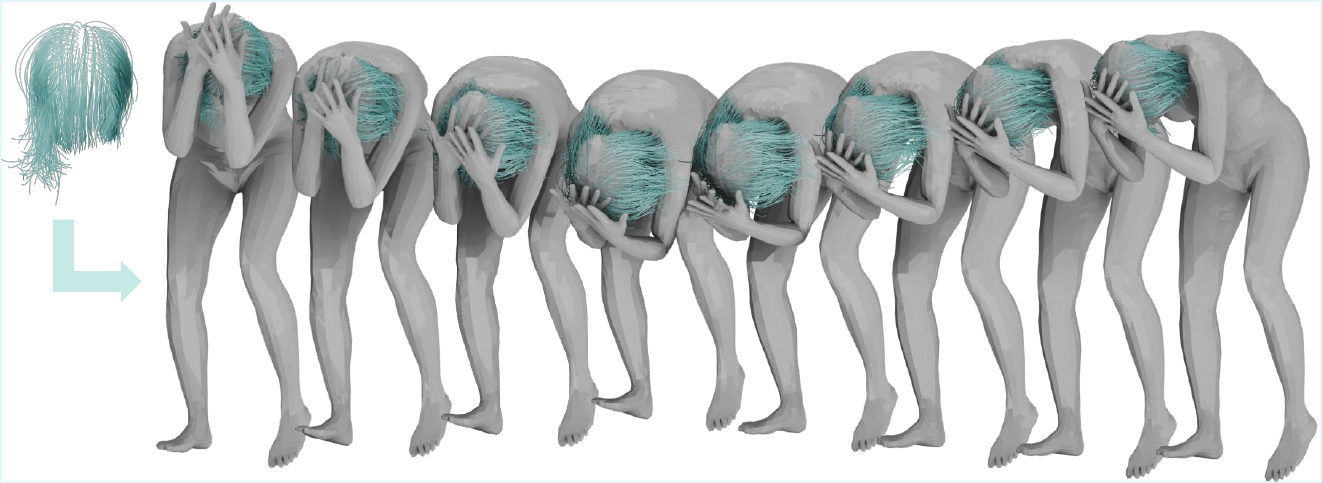}
\vspace{-4mm}
\caption{Per-frame inference using our static network for an unseen hairstyle. Our network correctly predicts the back hair splitting, while maintaining penetration-free drapes while allowing hair to slide along the arm.}
\label{fig:static-only}
\end{figure*}

We demonstrate HairFormer's robust performance in static and dynamic hair simulation through a range of experiments using various \emph{unseen} hairstyles and pose sequences. Our static network demonstrates strong capabilities in predicting physically plausible hair drapes. As Figure \ref{fig:cnn} illustrates, it accurately predicts the direction of gravity for various head rotations, effectively resolves penetrations (marked in red) for challenging unseen long hairstyles (also in Figure \ref{fig:penetration}), and, as shown in Figure \ref{fig:static-only}, precisely renders the natural splitting of the back hair when the posed figure bends downward.

\begin{figure*}
\vspace{-2mm}
\centering
\includegraphics[width=0.84\textwidth]{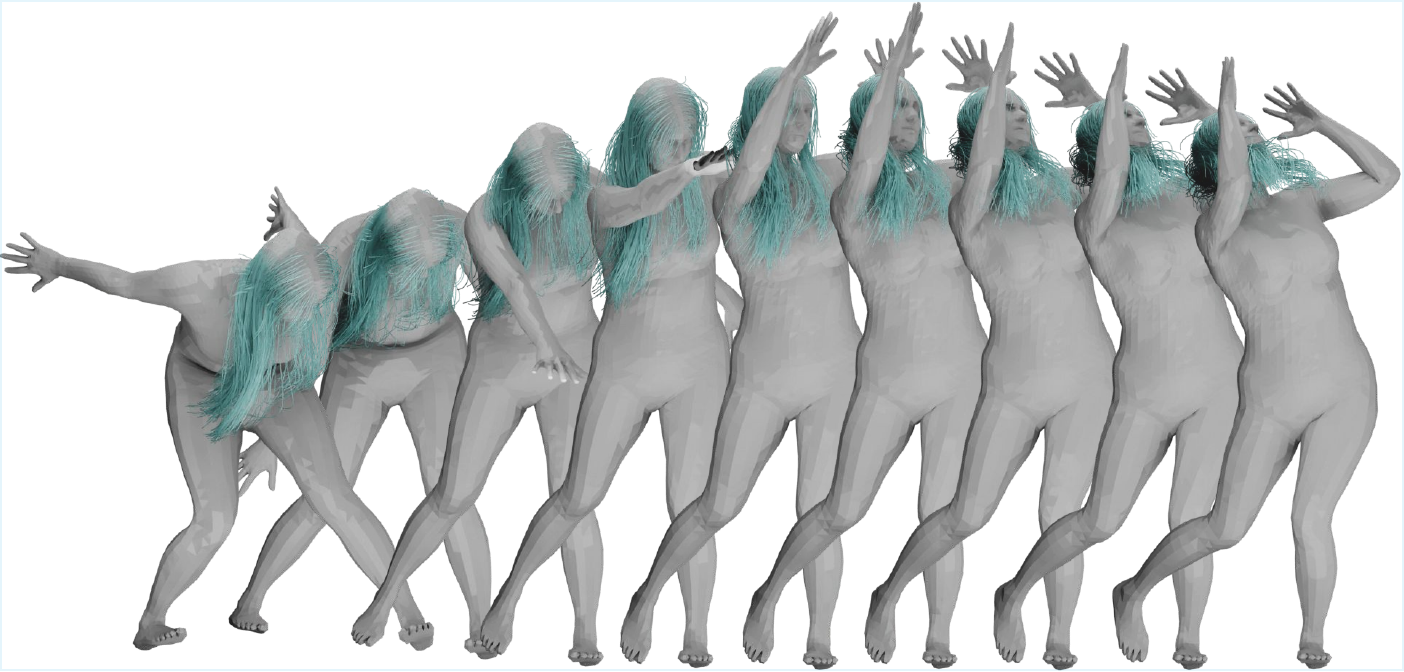}
\includegraphics[width=0.84\textwidth]{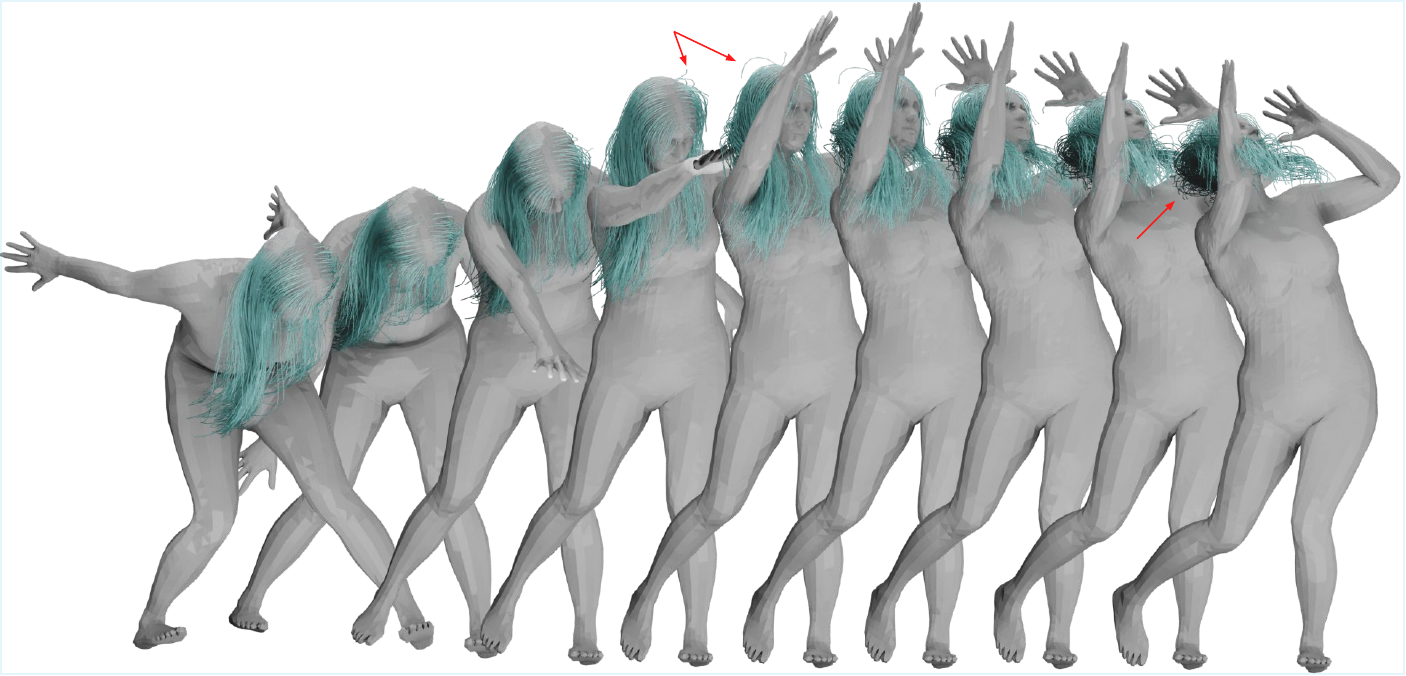}
\caption{Top: Static per-frame inference. Bottom: Dynamic inference for entire sequence. Red arrows indicate regions with significant differences.}
\label{fig:dynamic-static}
\end{figure*}

Beyond static drapes, our dynamic network further enhances realism by capturing expressive dynamic motions. Figure \ref{fig:dynamic-static} shows that even without specific fine-tuning, the dynamic network generates realistic hair flying and "antenna hair" effects during fast motions, which is unachievable by static methods. This inherent ability to produce rich, expressive dynamics is then further elevated for extreme scenarios. Figure \ref{fig:whiplash} shows that with only minor fine-tuning of a few hundred iterations, our dynamic network can accurately infer complex whiplash motions, where hair exhibits characteristic follow-through after a quick head rotation before recovering to its quasi-static draped shape.

\subsection{Ablation Studies}
\label{sec:ablation}
\subsubsection{Root and Auxiliary Loss Functions}
\begin{figure}[H]
\vspace{-12mm}
\centering
\includegraphics[width=0.98\columnwidth]{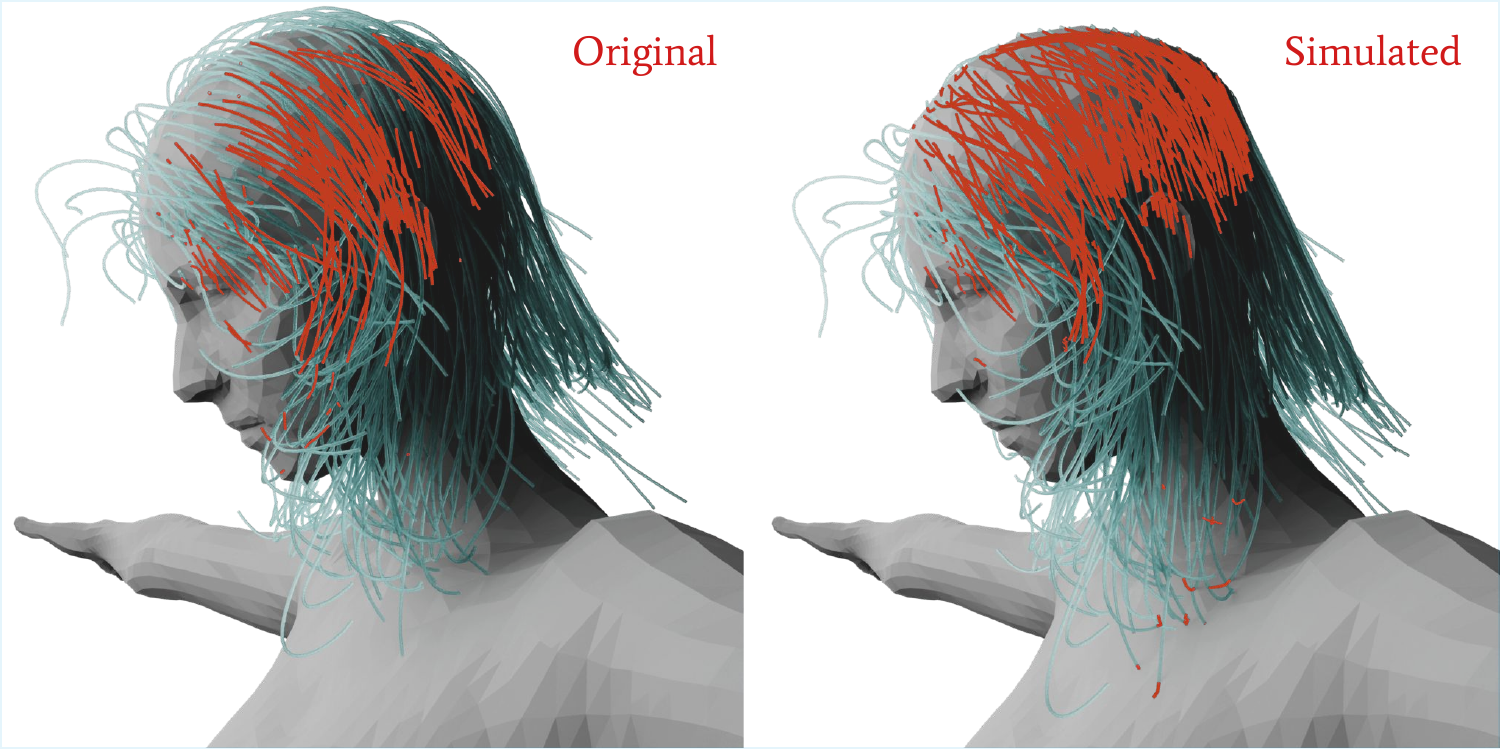}

\includegraphics[width=0.98\columnwidth]{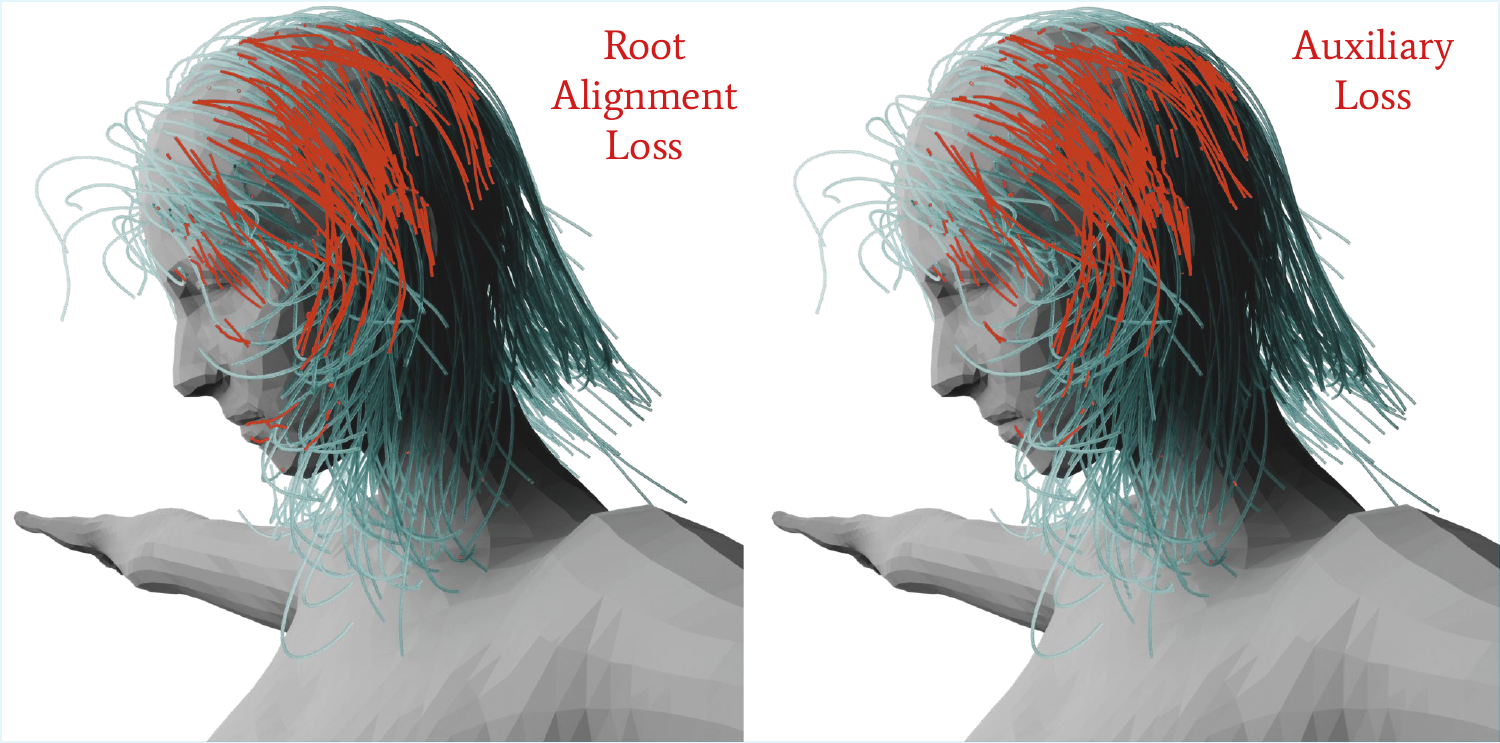}
\caption{Ablation study on loss functions: Rigidly-posed hair (top left), Adam-optimized hair without either loss, with $\mathcal{L}_{root}^{(t)}$ (bottom left), with $\mathcal{L}_{aux}^{(t)}$ (bottom right). The losses resolve penetrations toward the hair roots.}
\label{fig:ablation-loss}
\end{figure}

We evaluate the effect of our loss functions in preventing hair sagging. Figure~\ref{fig:ablation-loss} shows that the root alignment loss $\mathcal{L}_{root}^{(t)}$ prevents root penetration, while the auxiliary loss $\mathcal{L}_{aux}^{(t)}$ captures pose-induced physics and maintains drape. Both are crucial for high-fidelity static hair.


\subsubsection{Dynamic Network Modules for Feature Fusion}
\begin{figure}[H]
\vspace{-5mm}
\centering
\includegraphics[width=0.95\columnwidth]{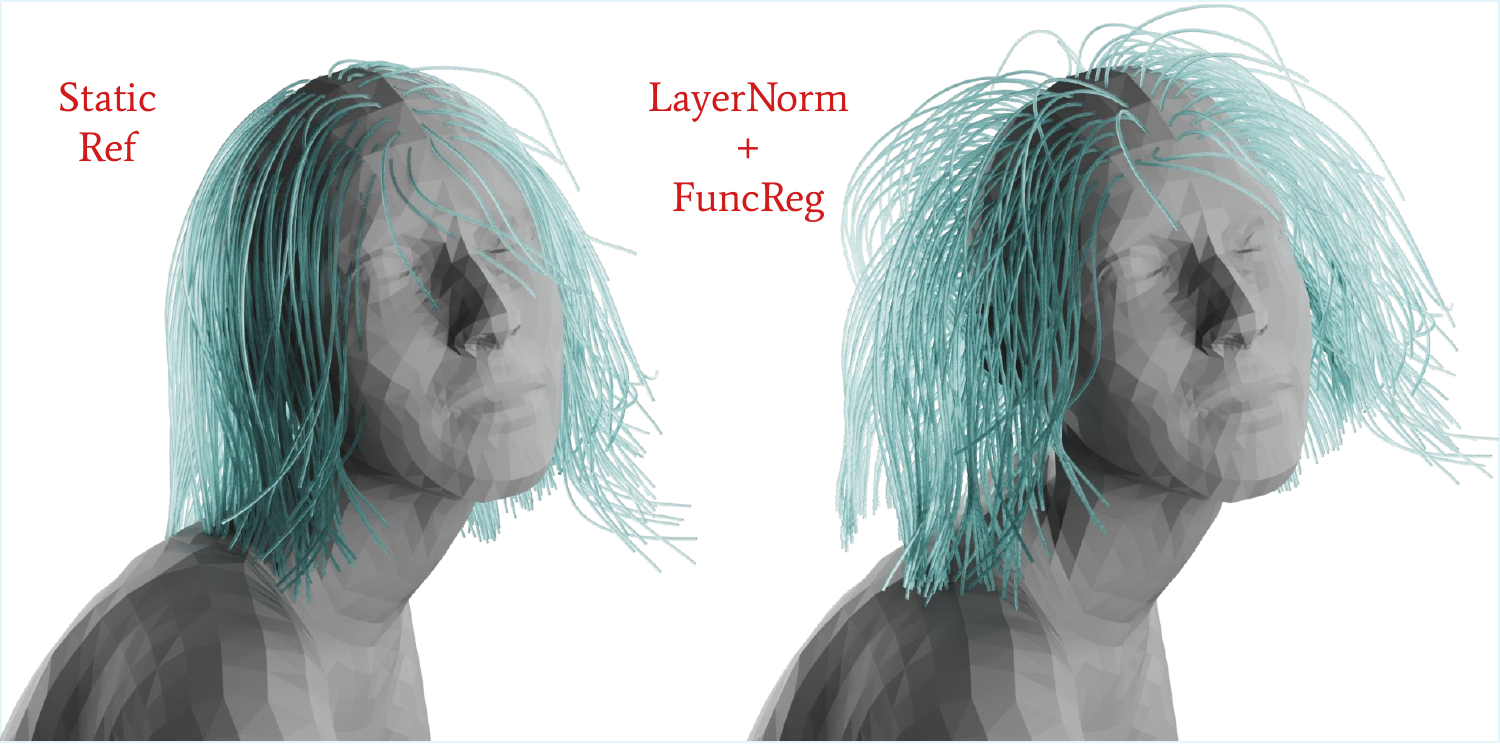}\vspace{1mm}
\includegraphics[width=0.95\columnwidth]{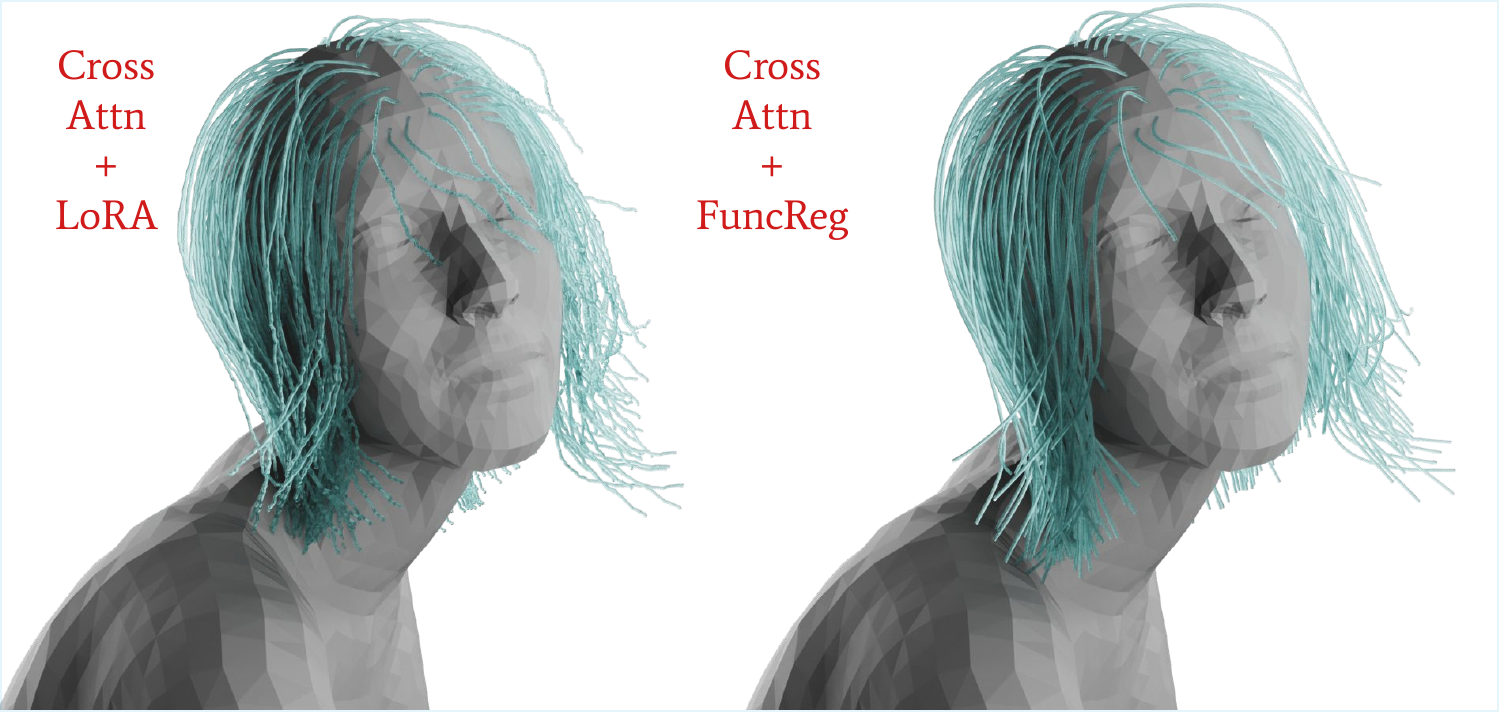}
\vspace{-3mm}
\caption{Ablation study on dynamic network modules: Static (top left), LayerNorm with $\mathcal{L}_\mathrm{func\_reg}^{(t)}$ (top right), cross-attention with LoRA (bottom left), cross-attention with $\mathcal{L}_\mathrm{func\_reg}^{(t)}$ (bottom right, our method).}
\label{fig:ablation-dynamic}
\end{figure}

We evaluate three strategies for fusing static ($\bm{H}_\mathrm{static}^{(t)}$) and dynamic ($\bm{C}_\mathrm{dyn}^{(t)}$) features, with the resulting dynamic drapes from a forward motion shown in~\cref{fig:ablation-dynamic} where the top left is the static reference.
\begin{enumerate}
  \item \emph{LayerNorm + FuncReg:} In this variant, instead of using cross-attention layers to fuse $\bm{H}_\mathrm{static}$ and $\bm{C}^{(t)}_\mathrm{dyn}$, we first use layer normalization to balance both features, and then directly concatenate them and feed the concatenated feature into an MLP prediction head to produce the deformation map result. We also adopt the functional regularization loss in \cref{eqn:funcreg_loss_t} in this variant. However, the ablation study results show that this variant yields exaggerated dynamics. This may be because the features $\bm{H}_\mathrm{static}$ and $\bm{C}^{(t)}_\mathrm{dyn}$ do not share the same feature space. Simply concatenating them is not effective in fusing the features, so the feature fusion result may be biased. This validates the effectiveness of our static-dynamic interaction blocks design, where the cross-attention mechanism is a powerful scheme in interacting features from different domains. This method results in exaggerated dynamics. 
 \item \emph{Cross Attn + LoRA:} In this variant, we adopt LoRA fine-tuning scheme~\cite{hu2022lora} rather than functional regularization during training the dynamic network to preserve static features. Specifically, we chose the LoRA rank and the alpha value to be 4 for both. However, it turns out that the LoRA fine-tuning cannot guarantee satisfactory convergence, making its result struggle to capture dynamics. 
 \item Our cross-attention layers with functional regularization scheme achieves both static feature preservation and the dynamic network's capability of capturing the dynamics. 
\end{enumerate}



\section{Limitations and Future Work}\label{sec:limitation}
\noindent\emph{Fixed Parameters:} Physical hair parameters are currently hardcoded, lacking support for user-defined inputs at inference. Future work could focus on developing an interface or mechanism to provide these parameters at runtime, offering greater flexibility and control.\vspace{1ex}

\noindent\emph{Missing Details:} High-frequency details like hair twist and fine curvature are not explicitly modeled, which could be extended by incorporating a residual network, such as the one used in PERM, for generating these intricate details.\vspace{1ex}

\begin{figure}[H]
\vspace{-5mm}
\centering
\includegraphics[width=0.98\columnwidth]{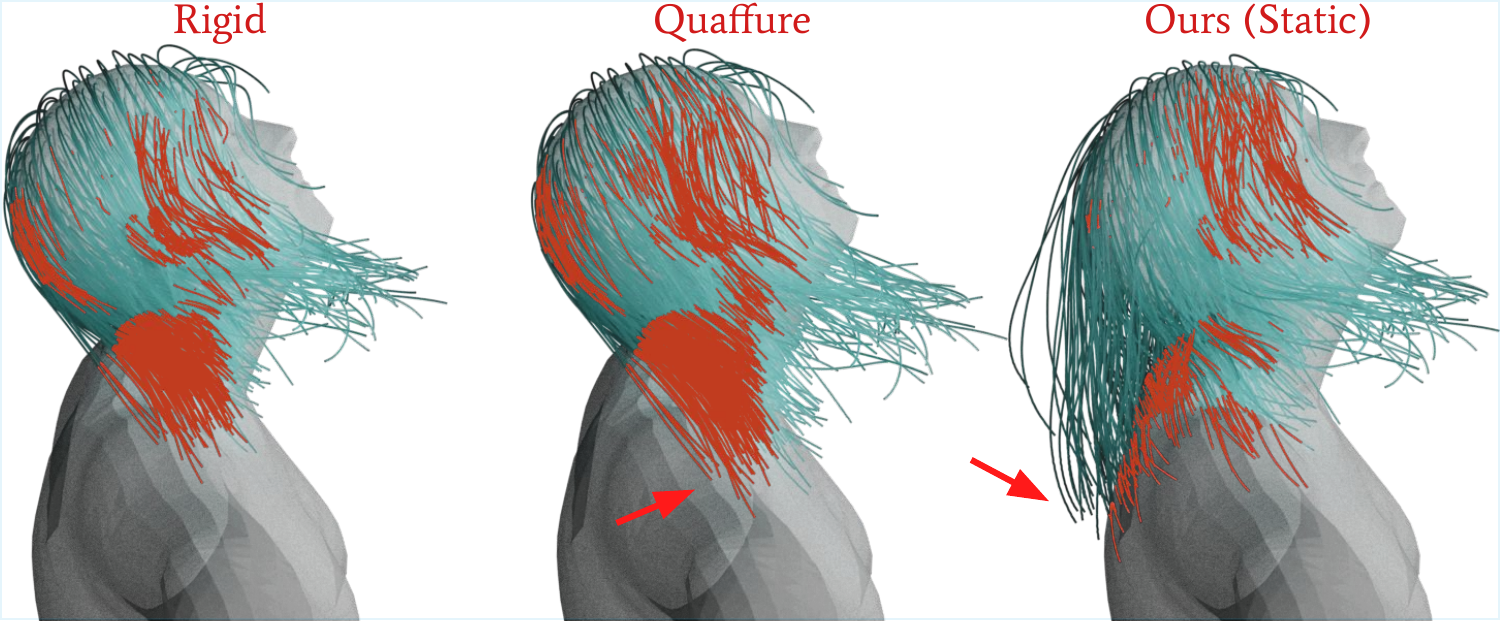}\vspace{-0.1mm}
\includegraphics[width=0.98\columnwidth]{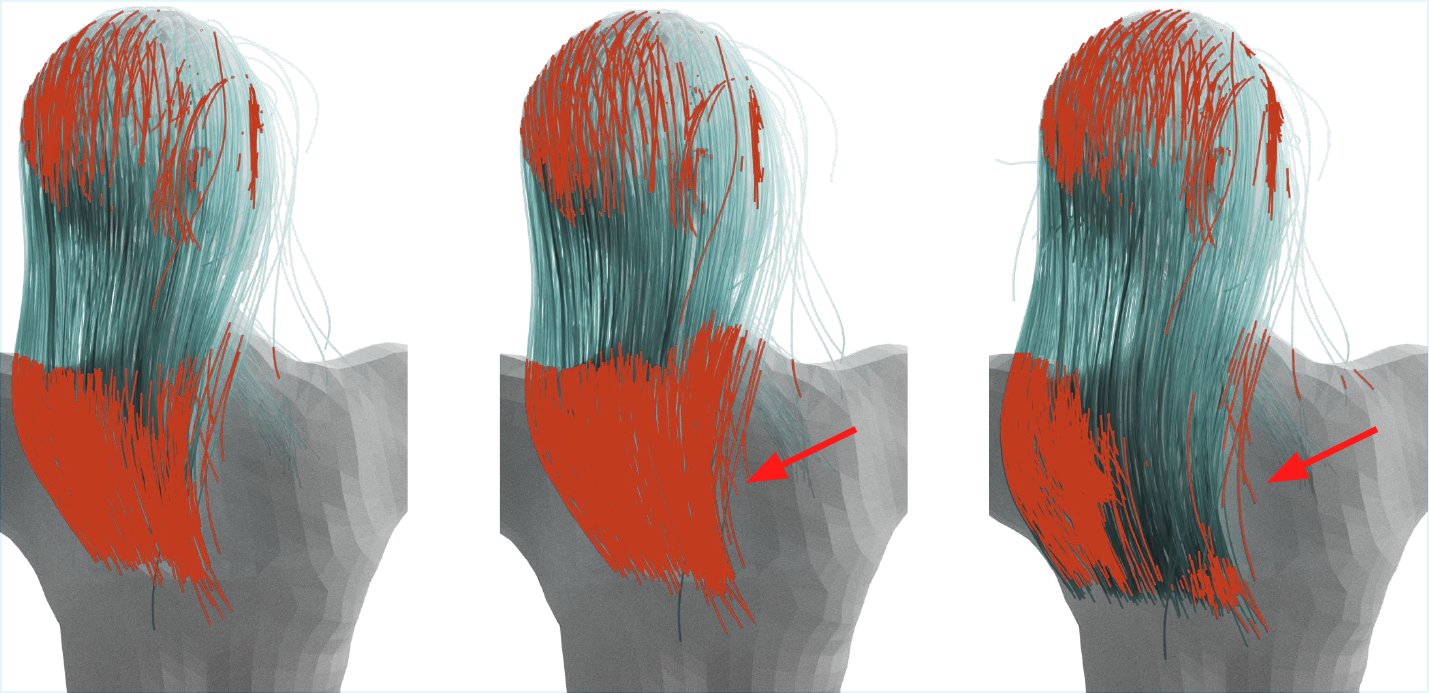}\vspace{-0.1mm}
\includegraphics[width=0.98\columnwidth]{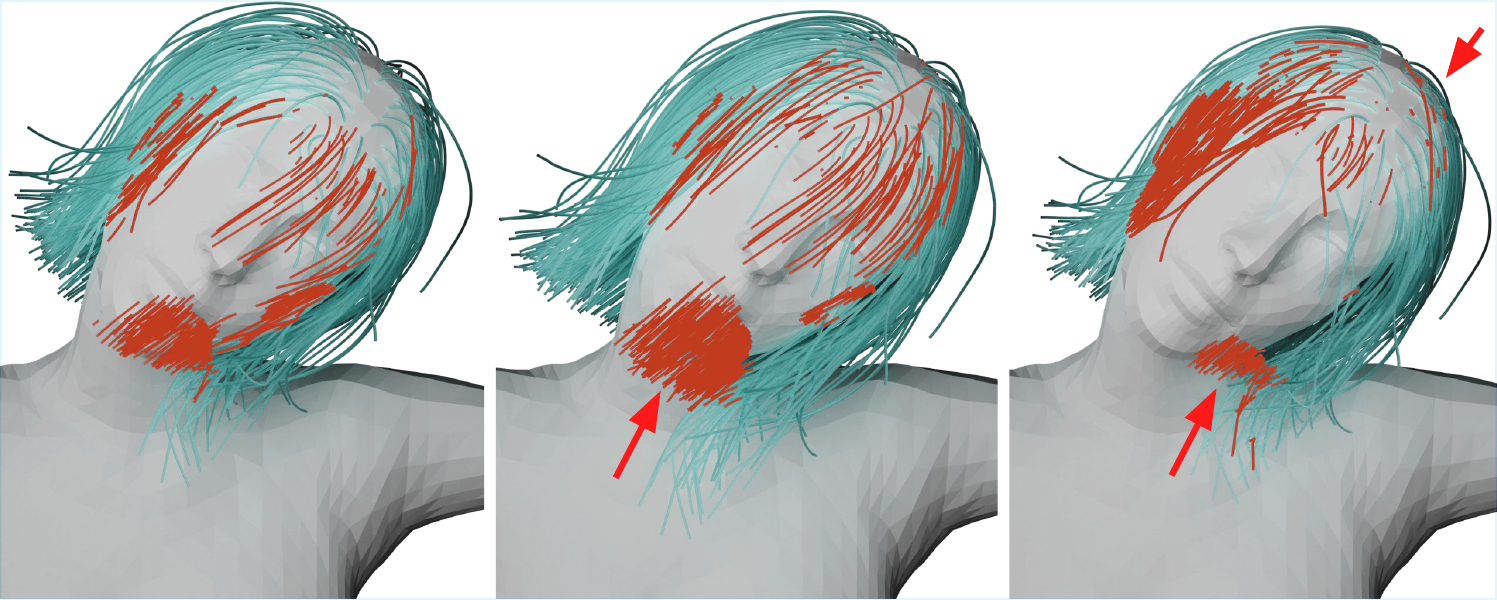}\vspace{-2mm}
\caption{Left to right: Rigidly-posed unseen hairstyles, Quaffure inference, HairFormer static network inference. 
For long hair (middle row), our method resolves penetrations into the back but cannot eliminate left shoulder penetrations, as discussed in the limitations.
}
\label{fig:cnn}
\end{figure}

\noindent\emph{Body Penetration:} 
Hair penetrations persist around complex body shapes (e.g., breast area, \cref{fig:cnn} middle) and unmodeled hand joints,
primarily due to insufficient training data. Given the generalizability of our model, a larger and more varied data set should significantly mitigate these problems. Future work will also explore more robust collision detection strategies or new shape-aware network modules.\vspace{1ex}

\noindent\emph{Inferred Dynamics:} Future research should focus on explicitly modeling velocity and acceleration rather than inferring them from pose sequences, which would allow for variable animation speeds without retraining and enhance the accuracy of friction modeling through improved detection of contact points and relative velocities.
\section{Conclusion}
We introduce HairFormer, a novel two-stage neural solution for dynamic hair simulation, a problem not addressed by prior work that largely focused on static hair simulation. Using Transformer-based architectures, our method achieves significantly fast inference times, successfully predicts physics-based, quasi-static draped shapes, and generates expressive dynamic motions with complex secondary effects over full pose sequences. The use of physics-informed losses ensures high-fidelity and generalizable dynamic hair that performs robustly across various styles and resolves penetrations effectively.

\bibliographystyle{ACM-Reference-Format} {
\bibliography{paper}
}
\clearpage
\newpage

\begin{figure}[h!]
  \centering
  \vspace{-10mm}
  \newcounter{imgFrameNum}

\begin{animateinline}[controls={play,stop}, autoplay, loop, width=0.49\textwidth]{4} 
    \multiframe{12}{iFrame=1+1}{
        \ifthenelse{\iFrame < 10}{%
            \includegraphics[width=\linewidth]{Figures/whiplash-carousel/pitch_frame_000\iFrame_body_dynamic_hair_tubes_front.pdf}%
        }{%
            \includegraphics[width=\linewidth]{Figures/whiplash-carousel/pitch_frame_00\iFrame_body_dynamic_hair_tubes_front.pdf}%
        }%
    }
\end{animateinline}
  \caption{Animated sequence of whiplash frames. (Requires a JavaScript-enabled PDF viewer like Adobe Reader to play animation). Please refer to our video for the full sequence.}
  \label{fig:whiplash}
\end{figure}

\newpage

\end{document}